\documentclass[aps,prx,twocolumn,nofootinbib,superscriptaddress,reprint,floatfix]{revtex4-2}
\usepackage{amsmath}
\usepackage{bbm}
\usepackage{float}
\usepackage{amssymb}
\usepackage{ifpdf}
\usepackage{color}
\usepackage{soul}
\usepackage{xcolor}
\usepackage{graphicx,epsfig}
\usepackage{hyperref}
\usepackage{tikz}
\usepackage{enumitem}
\usepackage{rsfso}
\usepackage{afterpage}
\hypersetup{
    colorlinks = True,
    linkcolor = {cyan},
    linkbordercolor = {cyan},
    citecolor = {blue},
    citebordercolor = {blue},
    urlcolor = {blue},
}

\usepackage[utf8]{inputenc}
\usepackage{tikz}
\usepackage{pgfplots} 
\usepackage{graphicx}
\usepackage[inkscapelatex=false]{svg}
\usetikzlibrary{shapes.geometric, arrows}

\tikzstyle{process1} = [rectangle, rounded corners, minimum width=3cm, minimum height=1.4cm,text centered, draw=black, fill=green!30, align=center]
\tikzstyle{process2} = [rectangle,rounded corners, minimum width=3cm, minimum height=1.4cm, text centered, draw=black, fill=blue!30, align=center]
\tikzstyle{arrow} = [thick,->,>=stealth]

\usepackage[normalem]{ulem}

\pgfplotsset{compat=1.16} 
\begin{document}

\title{Magnetically modified double slit based x-ray interferometry}

\author{S. Atkar}
\affiliation{Department of Physics, Indian Institute of Technology, Guwahati, Assam 781039, India}
\author{Z. Tumbleson}
\affiliation{Materials Sciences Division, Lawrence Berkeley National Laboratory, Berkeley, California 94720, USA}
\affiliation{Department of Physics, University of California, Santa Cruz, California 95064, USA}
\author{S. A. Morley}
\affiliation{Advanced Light Source, Lawrence Berkeley National Laboratory, Berkeley, California 94720, USA}
\author{N. Burdet}
\affiliation{Linac Coherent Light Source, SLAC National Accelerator Laboratory, Menlo Park, California 94720}
\author{A. Islegen-Wojdyla}
\affiliation{Advanced Light Source, Lawrence Berkeley National Laboratory, Berkeley, California 94720, USA}
\author{K. A. Goldberg}
\affiliation{Advanced Light Source, Lawrence Berkeley National Laboratory, Berkeley, California 94720, USA}
\author{A. Scholl}
\affiliation{Advanced Light Source, Lawrence Berkeley National Laboratory, Berkeley, California 94720, USA}
\author{S. A. Montoya}
\affiliation{Center for Magnetic Recording Research, University of California San Diego, La Jolla, California, USA}
\author{Trinanjan Datta}
\affiliation{Department of Physics and Biophysics, Augusta University, Augusta, Georgia 30912, USA}
\author{S. Roy}
\email{sroy@lbl.gov}
\affiliation{Advanced Light Source, Lawrence Berkeley National Laboratory, Berkeley, California 94720, USA}
\affiliation{Department of Physics, University of California, Santa Cruz, California 95064, USA}
 \begin{abstract}

We demonstrate an experimental approach to determine magneto-optical effects which combines x-ray magnetic circular dichroism (XMCD) with x-ray interferometry, based on the concepts of Young's canonical double slit. By covering one of two slits with a magnetic thin film and employing XMCD, we show that it is possible to determine both the real and the imaginary parts of the complex refractive index by measuring the fringe shifts that occur due to a change in the sample magnetization. Our hybrid spectroscopic-interferometric methodology provides a means to probe changes in the magnetic refractive index in terms of the electron spin moment.

\end{abstract}

\maketitle
Fundamental understanding of x-ray-matter interaction is the key to several advanced characterization methods with wide applications in many scientific areas \cite{JensAlsen2011, Stohr2006,ChemReviewXraySpect,Sakdinawat,acsnano.Bio0c09563}. Development of these methods rely on accurate and quantitative measurements of optical parameters \cite{OPticalTiN,OticalKAS2022110904,lucariniKramers}. Depending on the photon energy, x-rays can be used to extract different properties of a material or a sample. Soft x-rays (100-2000 eV), in particular, are used as a spectroscopic probe as it covers binding energies of several important elements such as carbon, oxygen, nitrogen. In addition, by tuning the incident x-ray beam energy to a $L$ or $M$ edge of a transition metal or rare earth element (the resonant condition) where the spectroscopic response is maximal, it is possible to achieve enhanced sensitivity to orbital and spin properties \cite{GibbsPhysRevB.37.1779,Hillsp0084}. An important example of a resonance based spectroscopy technique is x-ray magnetic circular dichroism (XMCD), which is the differential absorption of x-rays with opposite helicities \cite{TholePhysRevLett.68.1943,KaoPhysRevLett.65.373}. Together with x-ray absorption spectroscopy (XAS) and XMCD, it is possible to determine the complex refractive index $n(\omega) = 1 - \delta(\omega) + i\beta(\omega)$, where $\delta$ and $\beta$ are the real and imaginary components, respectively, describe the dispersive and the absorptive aspects of the x-ray-photon-material interaction. XMCD provides element-specific information about the magnetic moment, and, in conjunction with sum rules \cite{TholePhysRevLett.68.1943, CarraPhysRevLett.70.694}, can be used to obtain element-specific spin and orbital magnetic moments of the material \cite{Stohr2006}. 

\begin{figure*}[ht]
    \centering
    \begin{minipage}{0.5\textwidth}
        \centering
        \includegraphics[width=1.2\linewidth, height = 0.7\linewidth]{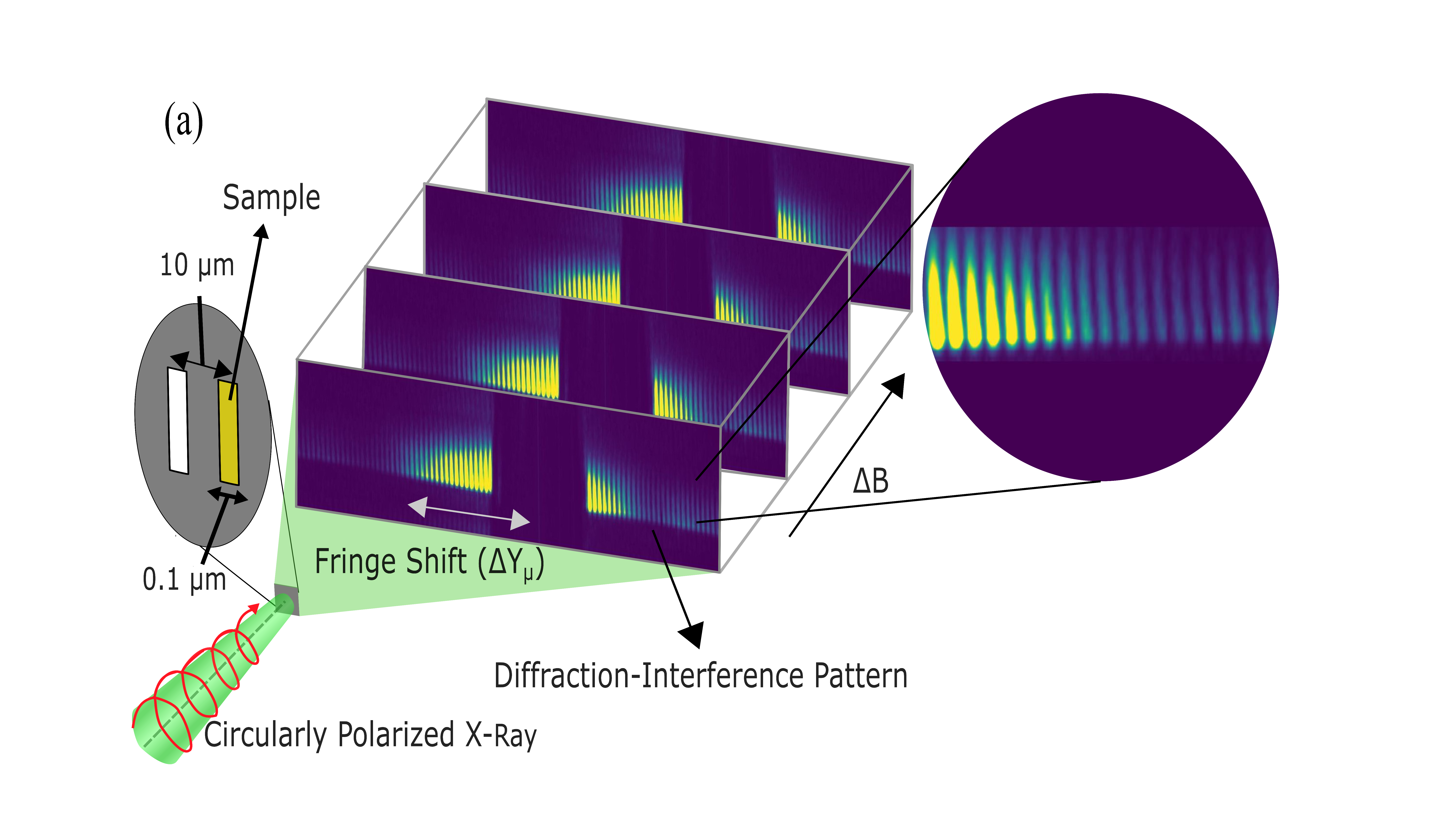}
        \vspace{22mm}
        \label{fig:subfig_a}
    \end{minipage}
    \hfill
    \begin{minipage}{0.49\textwidth}
        \centering
        \includegraphics[width=1.4\linewidth, height = 0.9\linewidth]{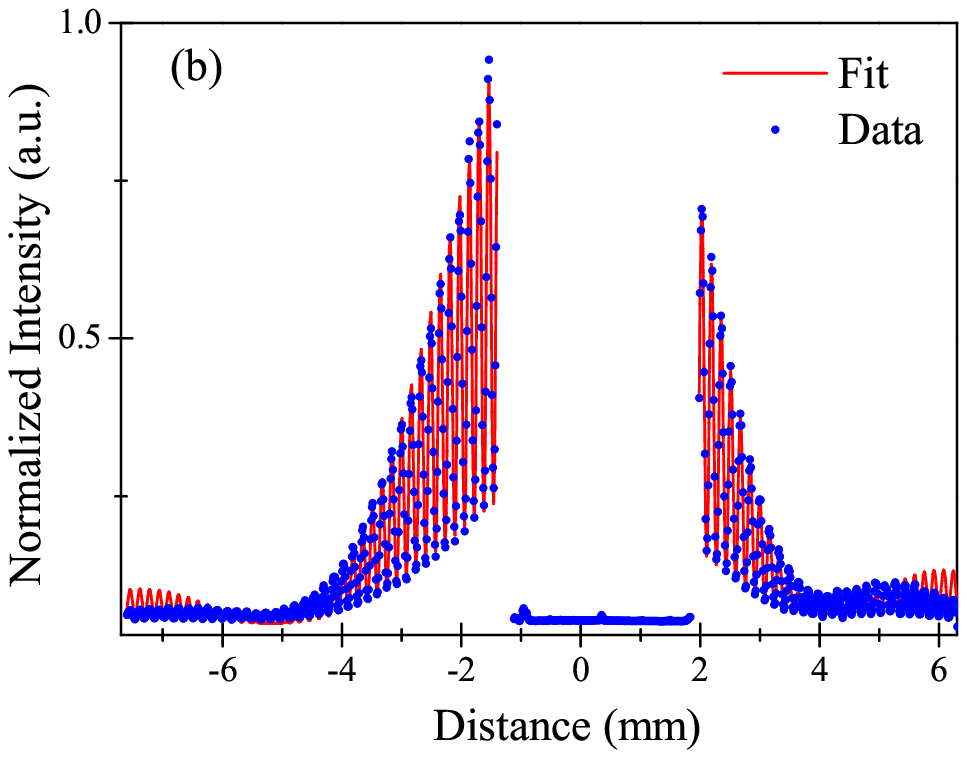}
        
        \label{fig:subfig_b}
    \end{minipage}
    \vspace*{-35mm}
    \caption{\small{(a) Schematic diagram of the soft x-ray Young's Double slit experimental set-up, with images recorded as a function of external OOP field at the Fe $L_{3}$ edge (707 eV). A zoomed-in view of the CCD images showcases the interference-diffraction pattern. (b) The resulting normalized intensity data (blue circles) overlapped with the resulting normalized intensity expression (red line) obtained from a non-linear least squares fit to the data. The missing intensity data at the center is due to the beamstop blocking part of the beam.}}
    \label{fig:mainfig}
\end{figure*}

In this Letter, we propose an innovative experimental set-up that integrates interferometry and XMCD to determine the magnetic refractive index. Utilizing a modified Young's double-slit arrangement (Fig.~\ref{fig:mainfig}a), we determine interference fringe pattern shifts induced by changes in the magnetic part of the refractive index of an Fe/Gd thin film heterostructure under an applied external magnetic field. We fabricated a specialized double slit with one open slit, and the other covered by the heterostructure film. The slits were illuminated by a coherent x-ray beam tuned to the resonant $L_{3}$ edge of Fe (707 eV). Light passing through the slits produces an interference fringe pattern on an x-ray charge-coupled device (CCD) camera downstream.

 The spin-dependent interaction of circularly polarized x-rays induce magnetic sensitivity in the double-slit diffraction pattern. In our measurements, we track the lateral shift of the fringe pattern for both the circularly \emph{plus} ($\sigma_{+}$) and the \emph{minus} ($\sigma_{-}$) polarized x-rays. By applying a varying external magnetic field and tracing hysteresis loops in the resultant pattern shift, we observe the coupling of optical properties to the applied field. This combination of interferometry with XMCD provides sensitivity to magnetization changes and enables us to identify the difference in the population of spin-up and spin-down electrons that gives rise to a net moment in the material.

At the resonant condition, the scattering factors become a complex quantity. The contribution to the scattering signal from the charge and magnetism can be determined. Consequently, it is possible to determine the charge and magnetic contribution to dispersion $\delta$ and $\beta$, as well as charge and magnetic thicknesses \cite{FreelandPhysRevB.60.R9923, RoyPhysRevLett.95.047201}. The interferometric method allows us to retrieve $\delta(\omega)$. Upon transmission, the phase (or \emph{optical path length}) of the light passing through the covered slit varies relative to the open slit. 
The optical path difference ${\Delta}l$ is related to the measured phase by $\phi = 2\pi\Delta l/\lambda$, where $\lambda$ is the x-ray wavelength. For a material thickness $t$, the path difference induced by transmission is related to the real-part of the refractive index by $(n(\omega) - 1)t = -\delta(\omega) t$. With prior knowledge of $\delta(\omega)$ (or thickness) and by analyzing the fringes of the resulting intensity interferogram as a function of the applied magnetic field, we can determine the \textit{magnetic} thickness, i.e., the thickness of the sample that produces magnetism and may or may not be equal to the actual thickness of the film of the sample (or \textit{magnetic} dispersion ($\delta(\omega)_{magnetic}$)). 

Experiments were conducted at the COSMIC-Scattering beamline (BL 7.0.1.1) of the Advanced Light Source at Lawrence Berkeley National Laboratory, using a canonical Young’s Double Slit configuration. (See Fig.~\ref{fig:mainfig}(a).) A high degree of transverse coherence in the incident x-ray beam was established with a 7 $\mu$m pinhole placed about 5 mm before the slits. The double slit consists of two slits, each nominally 100~nm wide, separated by $\approx$ 10~$\mu$m, center to center. The 100~nm $\times$ 60~$\mu$m slits were created by focused ion beam milling, producing one through-slit and one with the sample. The slit covered with our sample, consisted of a 30 nm-thick Fe/Gd multilayer. A 600-nm-thick layer of Au was deposited on its back-side to provide the slit mask with transmittance below $10^{-4}$. A Charge Coupled Device (CCD) camera, used to record the interference patterns, was placed 93 cm downstream of the slits.
Images were acquired with 400 ms exposure time, a frame size of 1024 $\times$ 1024 pixels, with $s_{\mathrm{pixel}}$ = 13.5~$\mu$m square pixels. In the data collection, 20 successive frames were averaged. A central beam-stop blocked the bright central region to prevent detector saturation and increase the dynamic range of the fringe pattern measurement. The incident beam energy was tuned to the Fe $L_{3}$ edge to achieve the resonant condition so as to get the magnetic sensitivity in the data. 

Two types of measurements were performed: (a) diffraction measurements as a function of incident beam energies, and (b) measurement as a function of the applied magnetic field. For the magnetic field experiments, and an out-of-plane external magnetic field was applied \textit{in-situ} within the field range of -1500 G to +1500 G, and back. All the measurements were performed using $\sigma_{+}$ and $\sigma_{-}$ incident x-ray light tuned to the Fe $L_{3}$ edge.  

We extract the fringe-pattern shifts ($\Delta Y$) from the recorded x-ray intensity images of the diffraction-interference pattern via a phase-correlation image registration algorithm \cite{988953}. This algorithm is resilient to noise and occlusions in the images of the recorded intensity patterns. Furthermore, sub-pixel precision is achieved by calculating a weighted centroid around the Fourier peak, eliminating the need for extensive upsampling of the Fourier transforms. In contrast, the first-order component of the Fourier Transform provides a deterministic assessment of 
$\Delta Y$ as interferences typically exhibit a single frequency. This approach has been previously employed for phase extraction from interferograms \cite{Goldberg:01}. 

The image registration algorithm uses the Fourier shift theorem \cite{988953}, which relates shifts in real space to phase differences in the frequency domain. By taking the inverse Fourier Transform of the cross-power spectrum, derived from the 2D Fourier Transform of the interferogram images, the position of the peak corresponding to the fringe shift ($\Delta Y$) is determined. The reliability of these values depend on the signal response of the cross-power spectrum, which is normalized to unity. The closer this response is to one, the more dependable the algorithm’s results are, indicating a distinct peak that accurately reflects the image shift.

The lateral fringe shift is $\Delta Y = \Delta\phi(L/2\pi)$, where $\Delta\phi$ and $L$ are the relative phase and the period of the interferogram, respectively. From the double-slit geometry, $L$ can be written as $L = \lambda s/d$ where $s$ is the slit-to-detector distance, and $d$ is the separation between the two slits. Substituting the relative phase in terms of the path length difference 
$\Delta\phi = 2\pi(1-\delta(\omega))t/\lambda$, we can combine these expressions to rewrite the lateral fringe shift as
\begin{equation}
\label{eq:phi}
    \Delta Y_{\mu} = \frac{(\delta_{\mu}(\omega) - \delta_{\mu o}(\omega))ts}{d},
\end{equation}
where $\delta_{\mu o}$ represents the zero field fringe shift pattern of the interferogram for $\sigma = \pm$ polarized light, which has an inherent phase accumulation due to the thickness of the film covering one of the slits. The complex component of the refractive index $\beta(\omega)$ accounts for x-ray absorption, and has no affect on the lateral fringe shift.
\begin{figure*}[t]  
  \centering
  \includegraphics[width=1.55\linewidth, height = 0.6\linewidth]{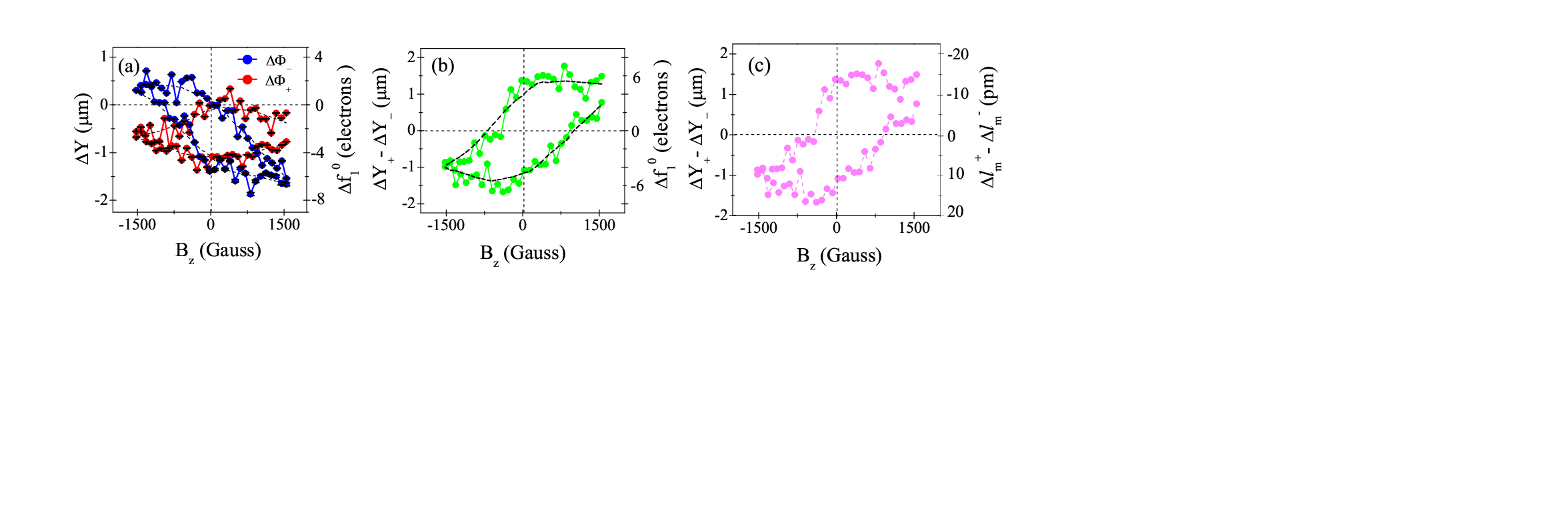}
  \vspace*{-60mm}
  \caption{Experimental fringe shift recorded through the magnetic hysteresis. (a) shows the registered fringe displacement for circularly polarized light. (b) shows the difference between the two curves. The axis on the right-hand side shows the corresponding change in the scattering factor values ($\Delta f^{0}_{1}$). (c) Shows the difference between the two curves scaled to the difference in the magnetic path length.}
  \label{HLoop}
\end{figure*}

Figure~\ref{fig:mainfig}(b) shows normalized 1D line-cuts extracted from the CCD images. We observe high contrast fringes with an intensity envelope as expected from double slit diffraction. The data were fitted with an intensity expression. 
\begin{align}
    I = I_{o}\frac{\sin^2\alpha}{\alpha^2}
  \cosh{\tilde{\beta}} 
    \left( \frac{\sin{\theta}}{\theta}\cos\tilde{\gamma} + 
    \cosh{\tilde{\beta}}\right) \label{eq:intenI}
\end{align}
where $\tilde{\beta}=\frac{2\pi\beta t_{f}}{\lambda}$ and $\tilde{\gamma}=2\gamma - \frac{2\pi\delta t_{f}}{\lambda}$ . The above equation was 
derived using the concepts of Fraunhofer diffraction \cite{Born_Wolf_Bhatia_Clemmow_Gabor_Stokes_Taylor_Wayman_Wilcock_1999}. In this expression \( \alpha = \pi by/\lambda s \), where \( s \) is the slit-to-screen distance and \( b \) is the slit width; \( \gamma = \pi dy/\lambda s \), where \( d \) is the slit separation. The parameter $\theta$ is the envelope function that represents the beam's coherence and affects fringe visibility \cite{10.1119/1.5047438}. Keeping the parameters $\lambda$, $s$ and $t$ constant, we perform a non-linear least squares fitting of the intensity pattern.
Fig.~\ref{fig:mainfig}(b) shows how the fitted expression compares to the normalized intensity data points. From the initial fits, we determine the dimensions of the double slit grating such as the slit width $b$ and slit separation $d$ to be 101.17 nm and 10.12 $\mu$m respectively, which is consistent with the nominal values of 100 nm and 10 $\mu$m. Successfully verifying these slit parameters reinforces the reliability of the intensity data and the fitting expression. 

In Fig.~\ref{HLoop}(a), we show the hysteresis loop in the phase shift due to the field-induced phase change of the signal. Figure~\ref{HLoop}(a) shows the hysteresis curves from the fringe shift for the $\sigma_{+}$ ($\Delta Y_{+}$) and $\sigma_{-}$ ($\Delta Y_{-}$) beams, denoted by the blue and red curves, respectively. A striking difference is evident in the directionality of the fringe shifts between the $\Delta Y_{+}$ and $\Delta Y_{-}$ curves.  Both the curves exhibit a familiar hysteresis behavior, with the fringes shifting in opposite directions under similar field conditions, though to varying degrees. 

Figure~\ref{HLoop}(b) is the loop obtained by subtracting the two separate loops ($\Delta Y_{+} - \Delta Y_{-}$) giving us a signal proportional to the net magnetization. We overlay the data with smoothed curves, shown as dotted lines, generated using a Savitzky-Golay filter to highlight the overall trajectory of the fringe pattern \cite{Savitzky1964}. Applying Eq.~\eqref{eq:phi} we can say that for \(\Delta Y \sim 0.1\) $\mu$m, the corresponding magnetically induced changes in \(\Delta(\delta(\omega) \sim 5 \times 10^{-5}\). The difference \(\Delta Y_{+} - \Delta Y_{-}\) stabilizes at 1.5 $\mu$m.

\begin{figure}
    
    \centering\includegraphics[width=1.5\linewidth, height = 1.1\linewidth]{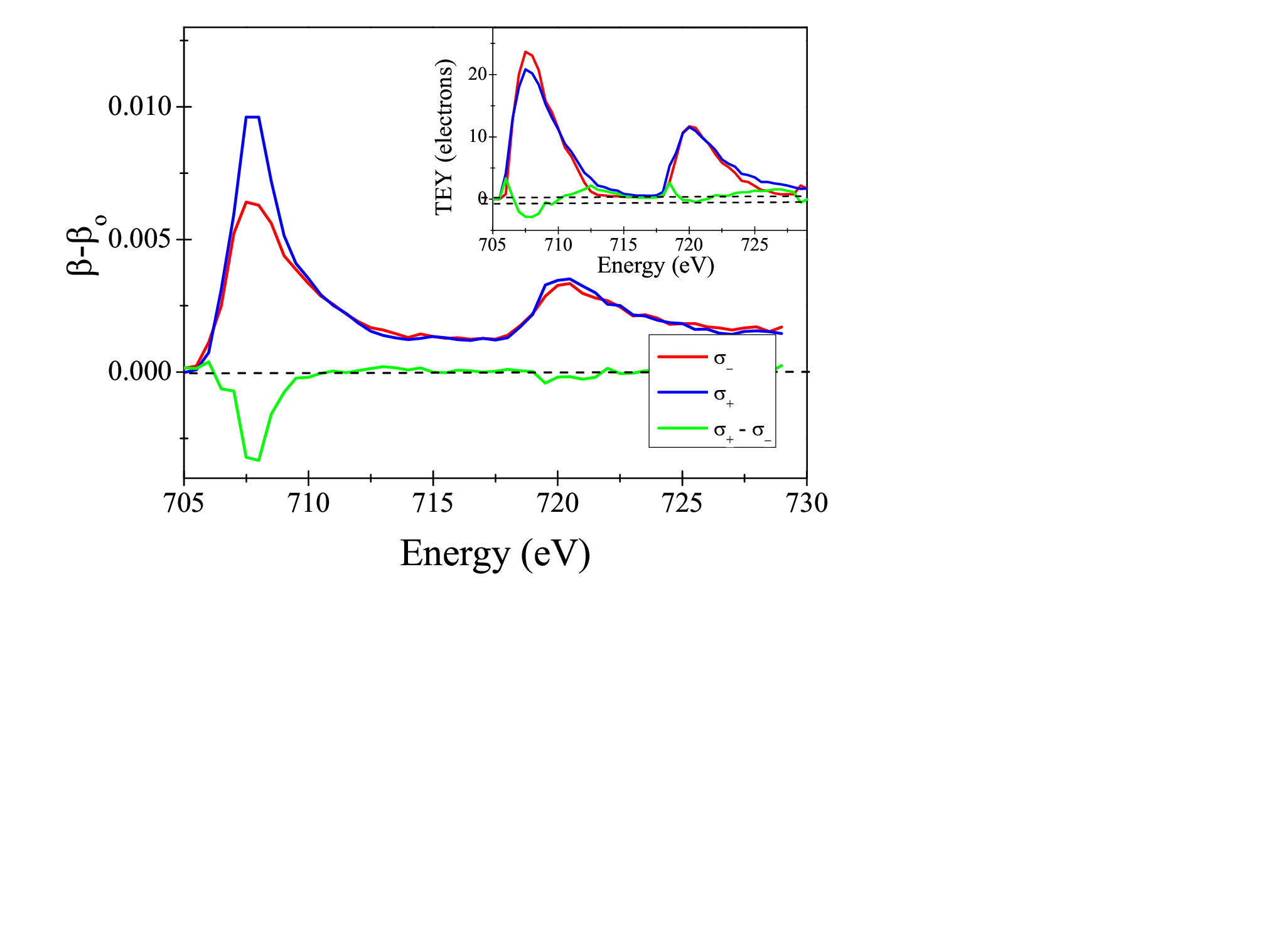}
    \hspace{-18mm}
    \vspace{-44mm}
    \caption{\small{The complex part of the refractive index vs Energy obtained from the fitting of the interferogram. Inset showcases the diode current values normalized to the total-electron-yield (TEY) as a measure of XMCD absorption.}}
    \label{fig:Escan}
\end{figure}

The expression for dispersion in the x-ray regime is given by $\delta(\omega) = n_{a}r_{e}\lambda^{2}f^{0}_{1}/2\pi$ where $n_{a}$ is the atomic number density of the species and $r_{e}$ is the electron radius \cite{Attwood_1999}. This expression, in conjunction with Eq.~\eqref{eq:phi}, enables us to relate the fringe shift change to the number of electrons that produces the refraction index change. The recorded fringe shifts in $\mu$m are scaled to $\delta(\omega) - \delta_{o}(\omega)$ using Eq.~\eqref{eq:phi} by substituting the known values of the Fe/Gd thickness $t$, detector-sample separation $s$, and slit separation $d$. Incorporating the previously mentioned expression for the x-ray dispersion, $\delta(\omega) - \delta_{o}(\omega)$ is scaled to represent the change in the number of electrons contributing to the magnetic moment $f^{0}_{1}(\sigma, \textbf{B}) - f^{0}_{1}(\sigma, 0) = \Delta f^{0}_{1}(\sigma)$. The atomic number density of the Fe/Gd sample is estimated by calculating the stoichiometric weighted mean of the elemental number densities using the expression $n_{a} = N_{A}\rho/M$ where $\rho$ is the mass density of the element, $M$ is the molar mass and $N_{A}$ is Avogadro's number This calculation yields an atomic number density of 8.53 $\times$ 10$^{28}$ atoms/m$^{3}$ enabling us to retrieve $\Delta f^{0}_{1}$ from the fringe shift data. Furthermore, to plot the change in the magnetic path length $l_{m}$ as shown in Fig.~\ref{HLoop}(c) we employ the expression $l_{m} = (1-\delta(\textbf{m}))t$ resulting in a direct relation between the relative path length difference and the relative change in the dispersion $\Delta l_{m} = -\Delta \delta t$. We note that the values of $\delta_{o}(\omega)$ obtained is consistent with published values in the literature \cite{KimPhysRevB.62.12216}. 

To illustrate the capability of the double slit to isolate the Fe $L_{3}$ edge at approximately 707 eV, we performed energy scans and extracted the $\beta(\omega)$ parameter from the resulting interferograms using the method described earlier, shown in Fig.~\ref{fig:Escan}. The two sharp peaks correspond to the Fe absorption edges and exhibit the XMCD effect, appearing as a differential absorption between the two helical beams. The peaks in the $\beta(\omega)$ fits are consistent with the peaks seen in the total-electron-yield (TEY) data (shown in Fig.~\ref{fig:Escan}(inset)). We note that the differential absorption is pronounced at the $L_{3}$ edge, but this is not the case at the $L_{2}$ edge. A similar lack of absorption contrast has previously been observed in Fe$_{3}$Gd$_{3}$O$_{12}$, where Gd atoms are antiferromagnetically coupled to the Fe moments, reducing the net magnetization at the $L_{2}$ energy \cite{RUDOLF1992109}.
\begin{figure}
    \centering    
    \includegraphics[width=1.45\linewidth]{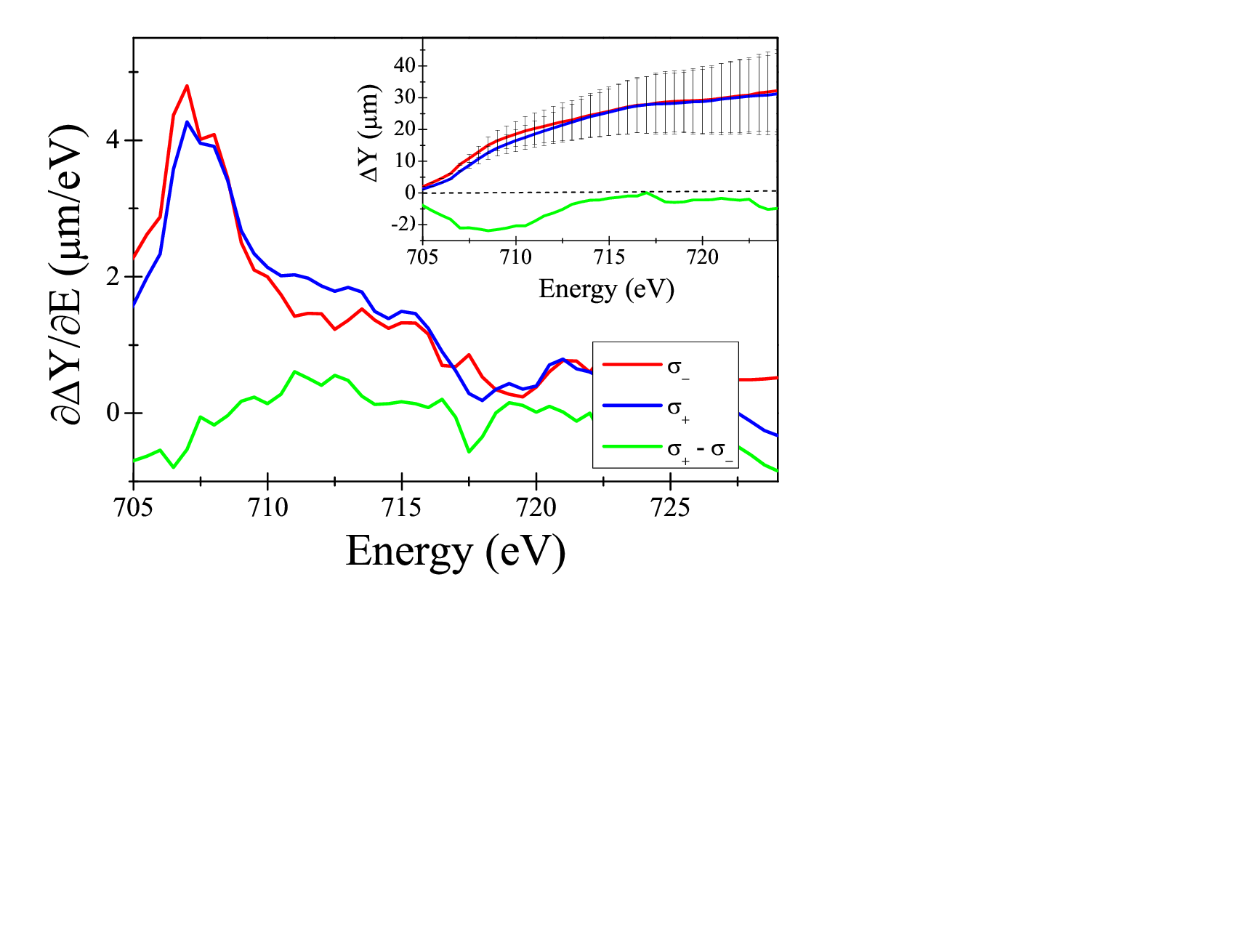}
    \vspace{-44mm}
    \caption{\small{The rate of fringe shift as a function of the beamline energy as measured by the image registration algorithm. The inset tracks the absolute shift for $\sigma_{+}$ and $\sigma_{-}$. The difference between them is scaled to enhance its features.}}
    \label{fig:RateShift}
\end{figure}

By tracking the fringe shifts as a function of the photon energy for $\sigma_{+}$ and $\sigma_{-}$, we can isolate the Fe $L_{3}$ edge. Fig.~\ref{fig:RateShift} displays the rate at which the fringes shift across the beamline energy while the inset tracks the absolute position of the fringes. The dominant peak in the $\partial\Delta Y/\partial E$ data corresponds to the Fe $L_{3}$ edge. The fringes move rapidly near the $L_{3}$ edge because the refraction index changes rapidly near the edge. The magnetic sensitivity is also enhanced at these energy ranges. However, fluctuations in the intensity pattern across the energy compromise the image registration algorithm's accuracy, which may explain difficulties in distinguishing the $L_{2}$ edge through the fringe shift of the energy scans.  

The sensitivity of the set-up is determined by the smallest detectable change in the dispersive part of the refractive index $\delta(\omega)$, derived from the quantitative measurement of the fringe shift $\Delta Y$. A conservative estimate of the  registration algorithm allows for a minimum reliable detectable shift $\Delta Y_{min}$ of 0.01 pixels \cite{988953}. Plugging this into a rearranged iteration of Eq~\eqref{eq:phi} yields $\Delta\delta_{min}(\omega) = \Delta Y_{min}s_{pixel}d/ts$. This implies that the pixel size or resolution of the CCD detector along with the slit-to-detector distance $s$  significantly impacts the sensitivity of the measurement. The separation between the two slits $d$ is constrained by the focus precision of the milling ion beam. Essentially, the number of pixels per fringe peak, which corresponds to the image resolution of a single fringe, determines the accuracy with which $\Delta\delta(\omega)$ can be isolated. Experimental adjustments, such as extending $s$ from 1 m to 5 m, combined with enhanced resolution, can improve the sensitivity of $\Delta\delta(\omega)$ to the order of 10$^{-6}$.

In conclusion, we have presented an innovative and straightforward approach for accurately measuring the dispersion of a Fe/Gd film using a modified Young's double-slit setup. This was demonstrated at the Fe $L_{3}$ edge using the contrast between circularly polarized light to obtain the net magnetic contribution to the dispersion changes. The results are applicable to any coherent light source and can be further enhanced with alternative reconstruction schemes to improve precision. Increasing the detector-sample separation can allow for the registration of even smaller changes. Using a revised double-slit setup, where only the top half of one slit is covered by the thin film, can generate two stacked interferograms, providing a reference fringe pattern to more accurately track the fringe shifts. Our work also opens the door to exploring quantum materials like ferroelectrics and multiferroics, where optical properties change under an applied electric field. Additionally, combining single-shot differential measurements from the double-slit setup with a pump-probe experiment could offer an avenue to investigate Floquet physics in quantum materials by capturing time-dependent changes in the refractive index.

The authors thank Prof. Eric. E. Fullerton of UC San Diego for important discussions and insight. T.D. thanks Tom Colbert for insightful discussions on optics. This work was supported in part by the US Department of Energy Office of Science, Office of Workforce Development for Teachers and Scientists (WDTS) under the Science Undergraduate Laboratory Internship (SULI) program. Work at the ALS, LBNL was supported by the Director, Office of Science, Office of Basic Energy Sciences, of the US DOE (Contract No. DE-AC02-05CH11231)

\vspace{100cm}

\section*{Supplementary Material}
\section{Derivation of Eq. (2) of the main text}
The geometrical arrangement of the modified Young's double slit set-up with one of the openings covered by Fe/Gd film is shown below. This set-up is identical to the one displayed in Fig.~1 of the main text. The spatial distances involved in the experimental set-up gives rise to far-field (Fraunhofer) diffraction.

\begin{figure*}[hbtp]
    \centering
    \includegraphics[width=0.4\linewidth]{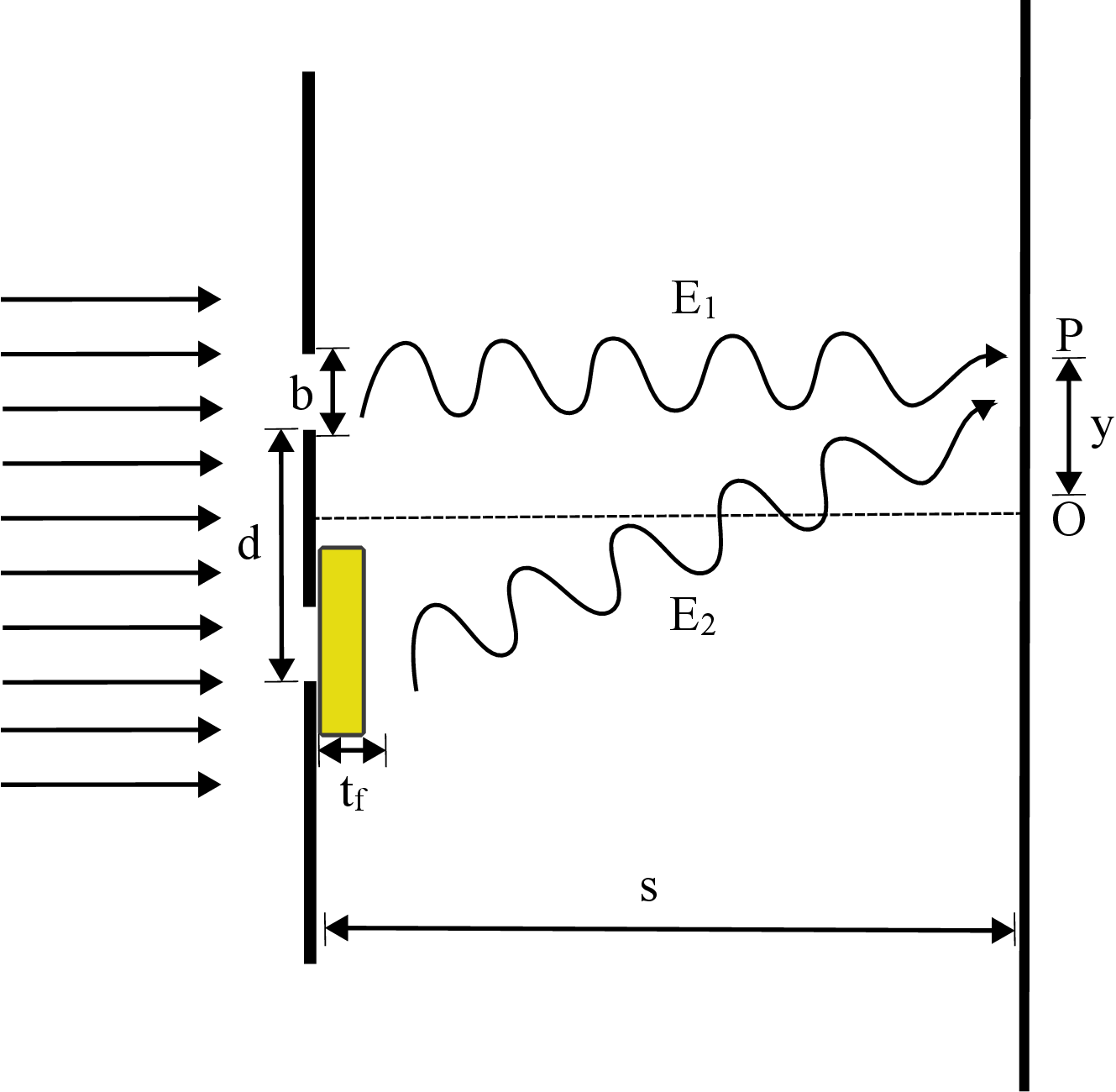}
    \caption{Fraunhofer diffraction of a plane electromagnetic wave (x-ray beam in our case) incident on the double slit grating, with one slit covered by a thin film. The slit width is $b$ and the slit separation is $d$. The screen, where the images are recorded using the CCD camera, is at a distance $s$. Considering the point $O$ to be the origin of the coordinate system, the distance $y$ gives the location of the fringes. The film thickness is denoted with $t_f$.}
    \label{fig:1s}
\end{figure*}

In the Fraunhofer regime, the expression of the time-dependent electric field $\mathbf{E}_1(t)$ for an arbitrarily polarized wave passing through a single slit of width $b$ (comparable to the slit separation $d$), with a phase difference of $\Xi$ between the electric field components is given by
\begin{equation}
\label{eq:e1}
    \mathbf{{E}}_{1}(t)  = E_{o}\frac{\sin\alpha}{\alpha}\left[\cos(\omega t - \alpha)\mathbf{\hat{x}} + {\alpha}\cos(\omega t - \alpha - \Xi)\mathbf{\hat{y}}\right],
\end{equation}
where $E_o$ represents the amplitude, $\omega$ is the frequency of the incident x-ray, and $\alpha = \frac{\pi by}{\lambda s}$. Here $s = 0.93$ m is the separation between the slits and screen, $b=100$ $\mu$m is the slit width, and $d =10$ $\mu$m is the separation between the slits. The second slit is covered by a magnetic film of thickness $t_{f}=30$ nm. Since $t_f \ll b$, all points of origin of the secondary wavelets emerging from the second slit are given by 
\begin{widetext}
\begin{align*}\label{eq:eq2}
    \mathbf{E}_{2}(t) = E_{o}\frac{\sin(\alpha)}{\alpha}\left[\cos(\omega t - \alpha - \Phi_{o})\mathbf{\hat{x}} \nonumber +\cos(\omega t - \alpha - \Xi  -\Phi_{o})\mathbf{\hat{y}}\right],
\end{align*}
\end{widetext}
where the phase difference $\Phi_{o}$ arises from the path difference due to the slit separation and the presence of the magnetic film over the second slit~\cite{Born_Wolf_Bhatia_Clemmow_Gabor_Stokes_Taylor_Wayman_Wilcock_1999}. This phase difference is given by
$\Phi_{o} = \frac{2\pi dy}{\lambda s} +  \frac{2\pi(n - 1)t_{f}}{\lambda}$, where $n = 1 - \delta +i\beta$ is the complex refractive index expression that describes the interaction of matter with x-rays. Here $\delta$ and $\beta$ represents the dispersive and the absorptive aspects of the refractive index. Considering an arbitrary point P at a distance $y$ from the center of the screen and using superposition, the net time-dependent electric field $\mathbf{E}_{net}(t)$ is given by 
\begin{widetext}
\begin{equation}
    \mathbf{E}_{net}(t)  = \mathbf{E}_{1}(t) + \mathbf{E}_{2}(t) = E_{o}\frac{\sin\alpha}{\alpha}\cos\frac{\Phi_{o}}{2}\left[\cos\left(\omega t - \alpha - \frac{\Phi_{o}}{2}\right)\mathbf{\hat{x}}+\cos\left(\omega t - \alpha - \Xi- \frac{\Phi_{o}}{2}\right)\mathbf{\hat{y}}\right].
\end{equation}
\end{widetext}
To obtain an expression for the intensity $I$, we consider the time-averaged integral of the modulus squared of the net electric field. We have $I = \langle |E_{x}(t)|^2+|E_{y}(t)|^2 \rangle_{T} = \frac{1}{T}\int_{0}^{T}(|E_{x}(t)|^2+|E_{y}(t)|^2)dt$ where $T = 2\pi/\omega$ is the time period of one cycle. Performing this integral yields the intensity expression as
\begin{widetext}
\begin{equation}
I = I_{o} \frac{\sin^{2}\alpha}{\alpha^2}  \cosh\left(\frac{2\pi\beta t_{f}}{\lambda}\right) \left[ \cos\left(\frac{2\pi d y}{\lambda s} - \frac{2\pi\delta t_{f}}{\lambda}\right) + \cosh\left(\frac{2\pi\beta t_{f}}{\lambda}\right) \right],
\end{equation}
\end{widetext}
where $I_{o} = \frac{E_{o}^{2}}{2}$. We can further simplify the expression by redefining variables $\tilde{\beta} = \frac{2\pi\beta t_{f}}{\lambda}$ and $\tilde{\gamma}=\frac{2\pi dy}{\lambda s} - \frac{2\pi\delta t_{f}}{\lambda}$. To incorporate the partially coherent nature of the light which in turn can affect fringe visibility, we add a coherence factor $\theta$ to the interference component \cite{10.1119/1.5047438}. This yields the final expression for the intensity [Eq. (2) reported in the main text] as 
\begin{align}
    I = I_{o}\frac{\sin^2\alpha}{\alpha^2}
  \cosh\tilde{\beta} 
    \left[ \frac{\sin{\theta}}{\theta}\cos\tilde{\gamma} + 
    \cosh\tilde{\beta} \right]. \label{eq:intenI}
\end{align}
The circular polarization of the beam affects the real and imaginary components of the refractive index $\delta$ and $\beta$. This in turn leads to the XMCD effect.  

\section{Fringe Shift ($\Delta Y$) and Absorption ($\beta$) Extraction}
\subsubsection{IIA.~$\Delta Y$ extraction}
The workflow for extracting the $\Delta Y$ parameter is summarized in Fig.~\ref{fig:2s}. In the first step, \textbf{Image compilation}, the CCD intensity images are systematically organized into a data set as a function of the varying experimental parameters, such as the applied magnetic field ($B$)
or beamline energy ($E$). In the next step, \textbf{Image Pre-processing}, we normalize by the beamline current to compensate for fluctuations in synchrotron intensity and apply a standard Savitzky-Golay filter to reduce noise in the intensity measurements without distorting the underlying signal \cite{Savitzky1964}. These steps ensure consistency and reliability in the data analysis before implementing the registration algorithm. 

\begin{figure*}[ht]
\centering
\begin{tikzpicture} [node distance=2cm]
    \node (box1) [process1] {Step 1:\\Image compilation \\};
    \node (box2) [process1, right of=box1, xshift=2cm] {Step 2:\\Image pre-processing \\};
    \node (box3) [process1, right of=box2, xshift=2cm] {Step 3:\\Shift registration \\algorithm};
    \node (box4) [process1, right of=box3, xshift=2cm] {Step 4:\\$\Delta Y$ plotting \\};

    \draw [arrow] (box1) -- (box2);
    \draw [arrow] (box2) -- (box3);
    \draw [arrow] (box3) -- (box4);

\end{tikzpicture}
\caption{Flowchart outlining the workflow to extract the fringe shifts ($\Delta Y$) from the recorded x-ray images. Step 1: Compile images into a dataset; Step 2: Pre-process images; Step 3: Register fringe shifts using the registration algorithm; and Step 4: Plot $\Delta Y$.}
\label{fig:2s}
\end{figure*}
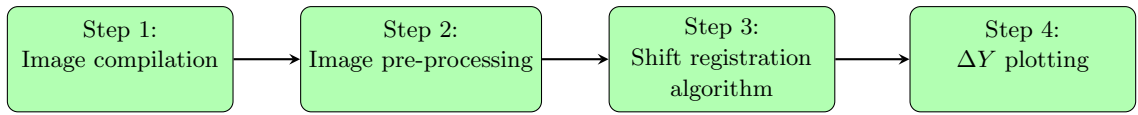
\begin{figure}[hbtp]
    \centering
    \includegraphics[width=0.8\columnwidth]{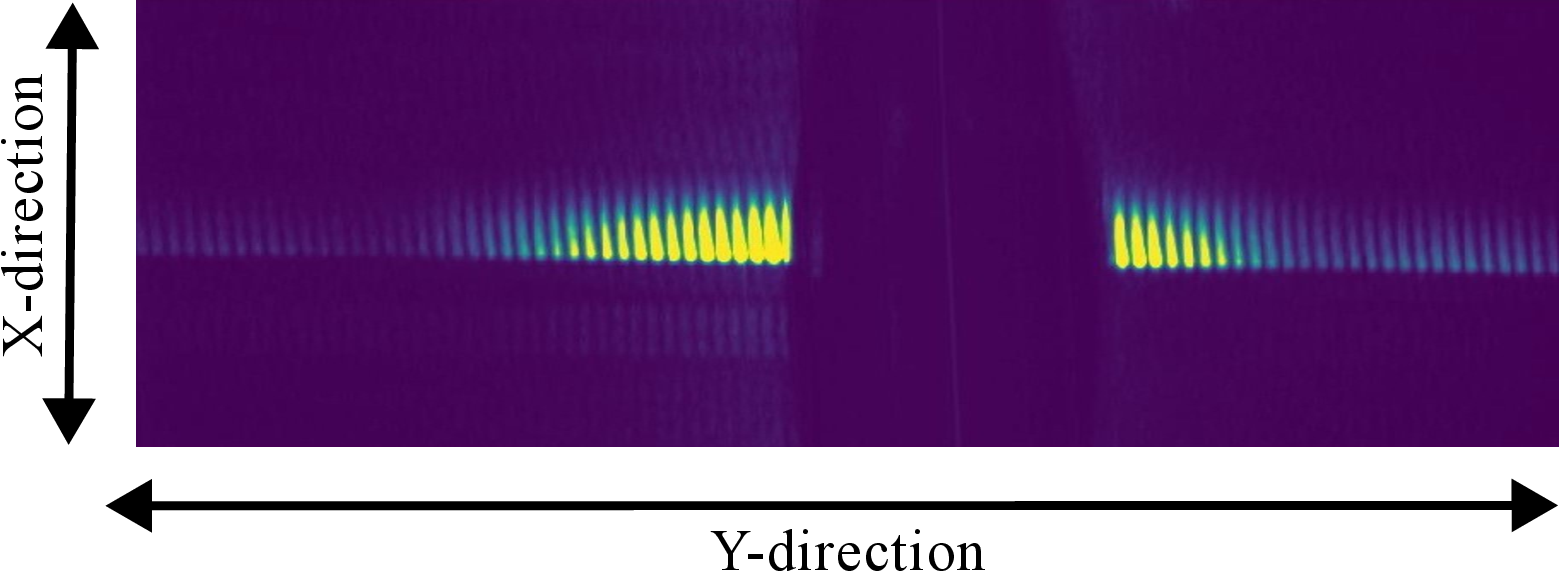}
    \caption{Intensity image of the fringe pattern generated by the double slit. The intensity values along the x-direction is averaged to yield a 1-D intensity pattern.
    }
    \label{fig:3s}
\end{figure}
Fringes located near the central region exhibit higher fringe visibility and strong signal strength. The image registration algorithm's accuracy is dependent on the signal-to-noise ratio (SNR), which can be characterized by the peak power percentage.  Therefore, 6–7 prominent fringes, close to the beamstop region, are carefully selected from the overall intensity pattern as the higher peak power ($\sim$ 90
\%) will yield more accurate results compared to fringes further away with a lower peak power \cite{988953}. To further refine the data, the intensity is averaged along the direction perpendicular to the fringe pattern (X-direction). Since we are exclusively interested in the Y-direction shifts in the fringes that are induced by the phase change in the beam, the X-direction averaging allows the algorithm to solely retrieve the $\Delta Y$ data. In step 3, \textbf{Shift Registration}, we use the first image in the series as a reference and the image registration algorithm tracks and quantifies the shift values across the subsequent fringe patterns. Finally, in step 4, \textbf{$\Delta Y$ plotting}, the extracted fringe shifts are plotted as a function of the varying experimental parameter (B or E).

\subsubsection{IIB.~$\beta$ extraction}

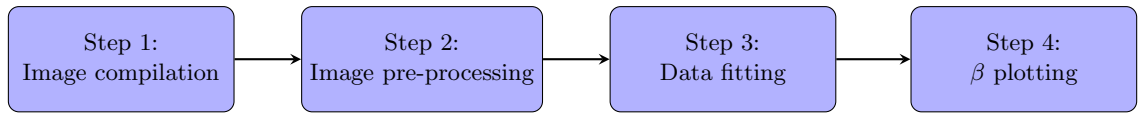
\begin{figure*}[ht]
\centering
\begin{tikzpicture} [node distance=2cm]
    \node (box1) [process2] {Step 1:\\Image compilation};
    \node (box2) [process2, right of=box1, xshift=2cm] {Step 2:\\Image pre-processing};
    \node (box3) [process2, right of=box2, xshift=2cm] {Step 3:\\Data fitting};
    \node (box4) [process2, right of=box3, xshift=2cm] {Step 4:\\$\beta$ plotting};

    \draw [arrow] (box1) -- (box2);
    \draw [arrow] (box2) -- (box3);
    \draw [arrow] (box3) -- (box4);

\end{tikzpicture}
\caption{Flowchart outlining the workflow to extract the absorption ($\beta$) from the recorded x-ray images. Step 1: Compile images into a dataset, Step 2: Pre-process images, Step 3: Extract the $\beta$ parameter by fitting to the data, and Step 4: Plot obtained $\beta$.}
\label{fig:4s}
\end{figure*}

To extract the $\beta$ parameter from the intensity images, the images are organized into a data set in step 1, \textbf{Image compilation}, and pre-processed in step 2, \textbf{Image pre-processing}, following a method similar to that described earlier. The general procedure is described in the flow chart in Fig.~\ref{fig:4s}. For step 3, \textbf{Data Fitting}, we use the least-squares fitting analysis and focus on one half of the intensity pattern, specifically excluding the beamstop region to avoid distortions caused by the blocked beam. To simplify the analysis, the intensity data is averaged along the direction perpendicular to the fringe pattern (X-direction), collapsing the two-dimensional intensity distribution into a one-dimensional curve, as shown in Fig.~\ref{fig:3s}. The obtained intensity curve can now be fit to Eq.~\eqref{eq:intenI} shown in the supplementary file [Eq.(2) of the main text]. Finally, in step 4, \textbf{$\beta$ plotting}, the obtained $\beta$ parameter is plotted as a function of the beamline energy.

\bibliography{GratingPaper/references}
 
\end{document}